\documentclass[sigconf]{acmart}
\usepackage{comment}
\usepackage[utf8]{inputenc} 
\usepackage[T1]{fontenc}    
\usepackage{hyperref}       
\usepackage{url}            
\usepackage{booktabs}       
\usepackage{amsfonts}       
\usepackage{nicefrac}       
\usepackage{microtype}      
\usepackage{lipsum}
\usepackage{amsthm}
\usepackage{subfigure}

\copyrightyear{2024}
\acmYear{2024}
\setcopyright{none}

\begin{document}

\title{A Reinforcement Learning Based Backfilling Strategy for HPC Batch Jobs}

\author{Elliot Kolker-Hicks}
\email{ekolkerh@uncc.edu}
\affiliation{%
  \institution{University of North Carolina at Charlotte}
  \city{Charlotte}
  \state{NC}
  \country{USA}
}

\author{Di Zhang}
\email{dzhang16@uncc.edu}
\affiliation{%
  \institution{University of North Carolina at Charlotte}
  \city{Charlotte}
  \state{NC}
  \country{USA}
}

\author{Dong Dai}
\email{ddai@uncc.edu}
\affiliation{%
  \institution{University of North Carolina at Charlotte}
  \city{Charlotte}
  \state{NC}
  \country{USA}
}

\begin{abstract}
{\color{black}\textit{This paper was originally published in the Workshops of the International Conference on High Performance Computing, Networking, Storage, and Analysis (PMBS 2023). This version has been updated to address several issues identified after publication.}}
\vspace{1em}

\noindent High Performance Computing (HPC) systems are used across a wide range of disciplines for both large and complex computations. HPC systems often receive many thousands of computational tasks at a time, colloquially referred to as “jobs”. These jobs must then be scheduled as optimally as possible so they can be completed within a reasonable timeframe. HPC scheduling systems often employ a technique called “backfilling”, wherein low-priority jobs are scheduled earlier to use the available resources that are waiting for the pending high-priority jobs. 
To make it work, backfilling largely relies on job runtime to calculate the start time of the ready-to-schedule jobs and avoid delaying them. It is a common belief that better estimations of job runtime will lead to better backfilling and more effective scheduling. However, our experiments show a different conclusion: there is a missing trade-off between \textit{prediction accuracy} and \textit{backfilling opportunities}. To learn how to achieve the best trade-off, we believe reinforcement learning (RL) can be effectively leveraged. Reinforcement Learning relies on an “agent” which makes decisions from observing the environment, and gains rewards or punishments based on the quality of its decision-making. Based on this idea, we designed RLBackfilling, a reinforcement learning-based backfilling algorithm. We show how RLBackfilling can learn effective backfilling strategies via trial-and-error on existing job traces. \textcolor{black}{Our evaluation results show up to 59\% better scheduling performance (based on average bounded job slowdown) compared to EASY backfilling using user-provided job runtime and 30\% better performance compared with EASY using the ideal predicted job runtime (the actual job runtime).}
\end{abstract}

\maketitle

\section{Introduction}


High-Performance Computing (HPC) systems provide key computational power for many critical scientific applications, such as climate modeling, high-energy physics, and astronomy~\cite{xgc,montage,kurth2018exascale}. To use such systems, scientists submit computational tasks, known as batch `jobs'. Then a centralized HPC job scheduler manages and runs these jobs in order. The order is determined by selected features of jobs, such as job waiting time (First-Come-First-Serve, FCFS) or job length (Shortest-Job-First, SJF). Given the volume of jobs HPC systems need to handle, effective scheduling is crucial.


Regardless of how jobs are ordered, HPC schedulers typically employ a technique called `backfilling' to further improve system efficiency and utilization. Specifically, if the next selected job (e.g. based on FCFS or SJF) cannot run because sufficient resources are not available, the scheduler nevertheless continues to scan the queue and selects lower-priority and smaller jobs that may utilize the available resources to run. A potential problem with this is that the first selected job may be starved as subsequent jobs continually jump over it. So, backfilling requires that the low-priority jobs are only allowed to run ahead provided they do not delay previously queued jobs or at least the currently selected job. This approach was originally introduced by EASY, the first backfilling scheduler~\cite{lifka1995anl}, and has been extensively used in various HPC job schedulers today~\cite{slurmweb,nitzberg2004pbs}. In this paper, we will use \textit{EASYBackfilling} or \textit{EASY} to refer to the backfilling strategy proposed in the EASY scheduler.


Backfilling needs to know the runtime of jobs. First, it needs to know when the currently running jobs will finish and release enough resources for the selected, pending job to start. Based on that, it can calculate a \textit{Reservation Time} for the selected job. Second, before backfilling a queued job, it needs to know if the job will finish before the Reservation Time so that it does not delay the execution of the selected job. Therefore, most HPC job schedulers require users to provide a runtime estimate for all submitted jobs, known as \textit{Request Time} or \textit{Wall Time}. 
The schedulers then leverage the user-specified information as an upper bound to estimate the finish of jobs for backfilling. However, relying upon user-submitted request time for backfilling is problematic as users will likely overestimate job runtime. The reason is that if a job goes over its request time, it will be killed or canceled by the scheduler. To avoid their jobs being killed, users often over-request the wall time~\cite{bailey2005user,mu2001utilization}. However, using a `larger' wall time to conduct backfilling will likely hurt the overall performance as previous work has shown~\cite{tsafrir2007backfilling,gaussier2015backfilling}.

An intuitive solution to correct this issue is to predict the job runtime more accurately using methods like machine learning (ML). If the prediction is accurate, then the schedulers are expected to conduct backfilling more effectively. Several recent studies have shown improved scheduling performance with more accurate runtime prediction~\cite{tsafrir2007backfilling,gaussier2015backfilling}. A natural question is then: \textit{does more accurate runtime prediction always lead to better scheduling and should better accuracy be pursued via more complicated ML models?}

Surprisingly, our simple experiments show a counter-intuitive result: higher runtime prediction accuracy does not necessarily lead to better scheduling performance, and pursuing better predictive models alone may not be sufficient to improve scheduling.  

\begin{figure}[h]
\begin{center}
    \includegraphics[width=0.4\textwidth]{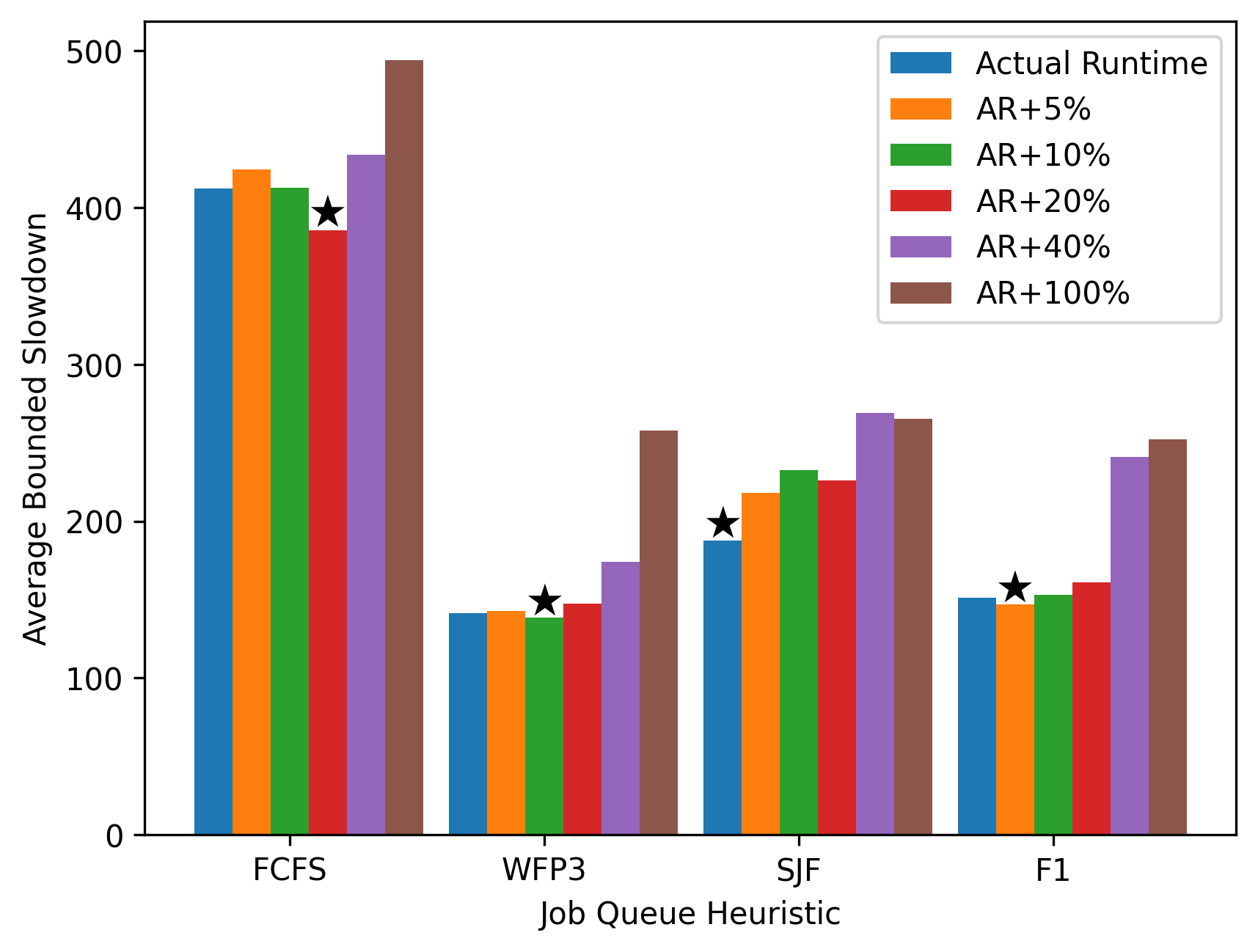}
	\caption{\textit{bsld} performance of different prediction accuracy on SDSC-SP2 using different scheduling algorithms.}
	\label{fig:fig1}
\end{center}
\end{figure}

To show that, we simulated scheduling the whole historical job trace (SDSC-SP2) collected from actual system~\cite{parallelworkloads}. In the trace, each job includes both user-submitted \textit{Request Time} (RT) and its \textit{Actual Runtime} (AR) after running. This allows us to experiment with job scheduling performance by using EASY backfilling with different job runtime prediction accuracy. Specifically, we used the actual runtime (AR) to mimic the scenario so that we can 100\% accurately predict the runtime. We then introduce random prediction errors in three cases (+5\%, +10\%, +20\%, +40\%, +100\%) to mimic imperfect runtime predictions. We calculate the average bounded job slowdown (\textit{bsld})~\cite{feitelson1998metrics} after scheduling the whole trace using several popular job schedulers with EASY backfilling enabled (FCFS, WFP3~\cite{tang2009fault}, SJF~\cite{pinedo2012scheduling}, and F1~\cite{Carastan-Santos2017ObtainingLearning}) and plotted the results in Figure~\ref{fig:fig1}. We selected \textit{bsld} as the performance metric as it has been used extensively in relevant studies and considered important to measure scheduling quality~\cite{Carastan-Santos2017ObtainingLearning,zhang2020rlscheduler,zhang2022SchedInspector} Here, slowdown means the ratio of job turnaround time over its execution time. The bounded slowdown measures job slowdown relative to given “interactive thresholds” (e.g., 10
seconds) to avoid overemphasizing short jobs.

From the results, we can easily observe that having better predictions on job runtime does not always lead to better job scheduling performance. There were several cases where noisy prediction led to the best overall performance. For instance, for the FCFS scheduling policy, using runtime prediction that has a +20\% noisy level leads to the best overall performance. Across all the experiments, only SJF could achieve the best performance using Actual Runtime during EASY backfilling. We conducted similar experiments on other job traces and got similar results. 

\begin{figure}[h]
\begin{center}
    \includegraphics[width=0.38\textwidth]{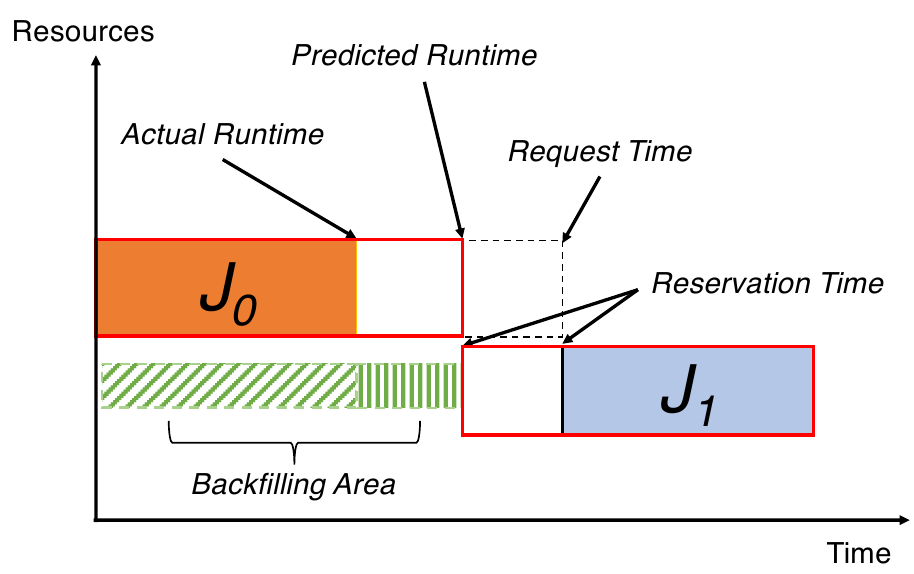}
	\caption{An illustrative example of trade-off regarding runtime prediction accuracy.}
	\label{fig:rational}
\end{center}
\end{figure}

The rationale behind this surprising observation can be illustrated in Figure~\ref{fig:rational}. Here, let's assume $J_0$ is a running job and $J_1$ is the selected next job and was waiting for the resources from $J_0$ to start. So, its earliest starting time (\textit{Reservation Time}) will be the end time of $J_0$. Since we do not know the exact end time of $J_0$, we use the \textit{Predicted Runtime}, which is typically some value smaller than \textit{Request Time} to conduct backfilling. As the figure shows, as the \textit{Predicted Runtime} moves closer to the \textit{Actual Runtime} (i.e., higher prediction accuracy), the \textit{Reservation Time} of $J_1$ also moves to the left, indicating an earlier start time and potentially better scheduling for it. But, at the same time, it also shortens the \textit{Backfilling Area} as the backfilled jobs need to finish before the \textit{Reservation Time} of $J_1$. This simply means fewer opportunities to run small jobs early, potentially hurting scheduling performance at the same time.

From this illustrative example, we can see there is a key trade-off that was largely missing in the existing study: using more accurate runtime to conduct backfilling will help the selected job to start earlier but also leave fewer opportunities for jobs being backfilled. Either extreme could lead to lower overall job scheduling performance as our experimental results have shown in Figure~\ref{fig:fig1}.

In this study, we propose RLBackfilling, a reinforcement learning (RL) based approach to directly learn efficient backfilling. Rather than pursuing accurate job runtime predictive models and then heuristically using the predicted runtime to decide backfilling, we train an RL agent to directly make decisions on which queued jobs should be backfilled when a backfilling opportunity arises. Such a decision depends on the estimated \textit{Reservation Time} of the selected job, the estimated runtime of queued jobs, and many other considerations. We expect the reinforcement learning agent would be able to learn such runtime estimations as well as the best trade-off between prediction accuracy and backfilling opportunities through trial and error. 



Our results showed that RLBackfilling was both flexible and effective. It can work with multiple base scheduling policies, such as FCFS, SJF, WFP3, and F1. Working with any of them, RLBackfilling outperforms EASYBackfilling using the \textit{Request Time} and up to 17x; outperforms EASYBackfilling using the \textit{Actual Runtime} up to 4.7x. Note that, these results are achieved by training the RL agent on the training dataset and testing it on separate testing workloads. 
In addition to this, we also show that RLBackfilling learns general rules of backfilling. The trained RL agent based on the FCFS scheduler outperforms other combinations of 'Scheduler'+EASYBackfilling. This further shows it was capable of easily adapting to a different HPC context. 

The remainder of this paper is organized as follows: In \S\ref{sec:background} we introduce the necessary background about backfilling and deep reinforcement learning. In \S\ref{sec:design} we present the proposed RLBackfilling and its key designs and optimizations. We present the main results (i.e. the RLBackfilling and its performances) in \S\ref{sec:eval}, and compare with related work in \S\ref{sec:related}. We conclude this paper and discuss the future work in \S\ref{sec:conclude}.

\section{Background}
\label{sec:background}

\subsection{HPC Scheduling and Backfilling}

This work discusses the scheduling and backfilling problem on HPC systems. We discuss its key aspects briefly.

\subsubsection{Job Attributes}
HPC batch jobs exhibit multiple attributes, including \textit{Submit Time}, \textit{Requested Nodes}, \textit{Requested Time}, and \textit{Wait Time}. Table~\ref{tab:attribute} provides an overview of commonly observed job attributes. For a comprehensive list of job attributes, refer to the Standard Workload Format
(SWF)~\cite{Feitelson2014ExperienceArchive}. 

\begin{table}[ht!]
    \caption{Description of job attributes.}
    \label{tab:attribute}
    \begin{center}
        \begin{tabular}{c|c|l}
        \hline
        \hline
        \textbf{Name} & \textbf{Symbol} & \textbf{Description}\\
        \midrule
        \textit{Submit Time}     & $s_t$ & job submission time \\
        \textit{Requested Nodes}    & $n_t$ &  the number of requested nodes.\\
        \textit{Requested Time} & $r_t$ & job's runtime estimation from users\\
        \textit{Wait Time} & $w_t$ & time job waits in the queue \\
        \hline
        \hline
        \end{tabular}
    \end{center}
\end{table}


\subsubsection{HPC Job Scheduling}
In HPC systems, users submit jobs to acquire resources to run their computational tasks. When submitting a job, users need to provide job size and runtime estimate before the scheduler will run it. Here, job size refers to the number of computing nodes in the HPC system. The runtime estimate (also known as Request Time/Wall Time) represents an upper bound of job execution. The scheduler will cancel or kill jobs that surpass their Request Time. 
The job scheduling in modern HPC systems typically uses a Base Scheduling Policy to decide which jobs in the queue should be scheduled next. The policy may use different job characteristics to assign priority. For example, \textit{First Come First Serve} (FCFS) uses the order of job arrivals to determine priority; \textit{Shortest Job First} (SJF) prioritizes jobs with shorter request time. Some machine learning-based schedulers, such as \textit{F1} \cite{Carastan-Santos2017ObtainingLearning}, consider multiple job features at the same time. 

\subsubsection{HPC Job Backfilling}
It is possible that the job selected by the Base Scheduling Policy can not run due to insufficient resources. Due to the priority limitation, all jobs need to wait until resources become available. This often leads to low system utilization.  
Backfilling is a scheduling technique used to run low-priority jobs ahead of schedule, as long as they do not delay the execution of the high-priority jobs. 
One of the most popular approaches to backfilling is EASY. EASY backfilling attempts to backfill when a job cannot be immediately scheduled (referred to as the relative job or rjob), and takes the opportunity to schedule as many jobs as possible so long as they do not interrupt the rjob that will run in the future when more resources are available. Backfilling has been proven effective in HPC systems. This study aims to explore more effective backfilling strategies.  


\subsection{Reinforcement Learning}
\subsubsection{Reinforcement Learning and Deep Neural Network Approaches}

Reinforcement Learning (RL) is a type of machine learning where an agent learns to make decisions based on information about its environment. It does this by receiving input on the current state of the environment (known as \textit{state space}), a list of possible actions it can take in the current environment (known as \textit{action space}), and a numeric output (known as \textit{reward}) based on how the choices it makes impacts the environment.

The agent initially does not know how its actions lead to rewards but learns over time through trial and error. It is then able to develop \textit{policies}, internal rules for decision-making based on the probability of taking actions in certain states. Depending on the problem RL is applied to, the \textit{state-space} (a space representing all potential system states) can become extremely large, making traditional RL approaches infeasible. To avoid this issue, a Deep Neural Network (DNN) can be used to predict action probabilities, a technique referred to as Deep Reinforcement Learning (DRL). In this study, we adopt the Proximal Policy Optimization (PPO)~\cite{schulman2017proximal} as our specific DRL technique. Not only is PPO a state-of-the-art method, but it is also simpler to implement and optimize. We opted for PPO over Deep-Q-Learning (DQL), another RL method that leverages DNNs, primarily because of the faster convergence assurances provided by policy-gradient methods compared to DQL\cite{convergencePolicyGrad}. Additionally, PPO stands as the default reinforcement learning algorithm at OpenAI due to its user-friendly nature and good performance~\cite{ppoweb}.

\subsubsection{Backfilling Defined as a Reinforcement-Learning Problem}

Intuitively, RL can be used for making backfilling decisions. We can define the current state of the scheduling environment (e.g., currently queued jobs) as the state space for the RL agent. 
Then, we can have the list of jobs available to backfill during a backfilling opportunity as the action space for the agent.
In this case, the reward for the agent will be based on a given performance metric after scheduling several jobs. Since Backfilling works after the base scheduling policy takes effect, we do need to consider how the RL agent would work with different scheduling policies such as SJF (shortest job first). We will further discuss the detailed design and implementation in the next section.



\section{Design and Implementation}
\label{sec:design}

Existing backfilling strategies essentially include two steps: 1) getting the job runtime estimations (either from user-submitted job script or predictions) and 2) using the job runtime estimations to heuristically select the best jobs to backfill. The key problem here is the second step is not aware of how noisy or inaccurate the job runtime estimations could be. Hence, the decisions could be largely sub-optimal.
RLBackfilling uses reinforcement learning to directly decide which job to backfill when backfilling opportunities arise during scheduling. It implicitly combines these two steps into a single one, which allows it to intelligently coordinate the runtime predictions and how to use the predicted runtime to backfill jobs.
In this section, we will describe the design and implementation of RLBackfilling.

\subsection{Overview}
RLBackfilling learns how to backfill jobs by simulating the scheduling process using public job traces. Each time, it is trained based on a given job scheduling policy, such as FCFS or SJF, and an optimization goal, such as minimizing average job waiting time or minimizing average job bounded slowdown. In this study, we focus on average bounded slowdown. We plan to explore other optimization goals in the future. 

\begin{figure}[h]
\begin{center}
    \includegraphics[width=0.48\textwidth]{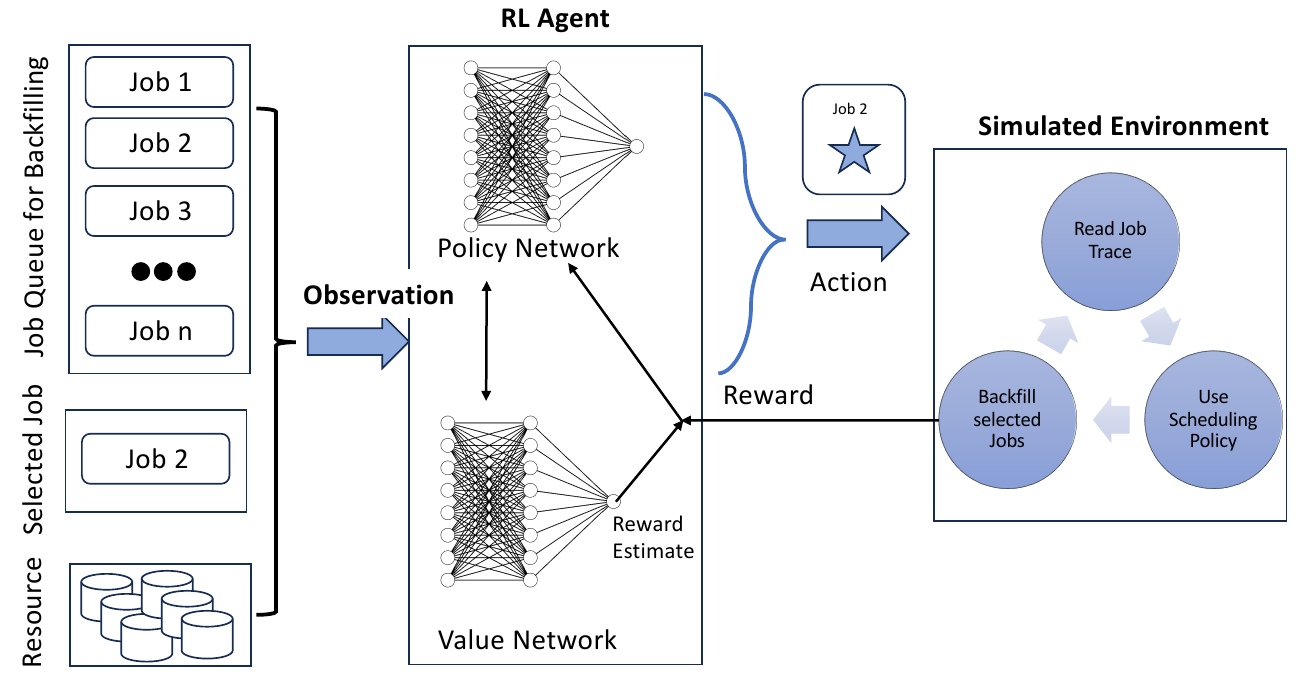}
	\caption{Overall architecture of RLBackfilling.}
	\label{fig:overall}
\end{center}
\end{figure}

Figure~\ref{fig:overall} shows the overall architecture of RLBackfilling. There are three main components to the design: a \textit{Simulated Environment}, a DNN-based \textit{RL Agent} that takes waiting queue and machine statuses as input to generate a backfilling action each time, and \textit{Observations} generated by the environment. At each time step, the agent uses the current observations to generate a backfilling action, which is applied to the simulated HPC environment to generate a \textit{Reward}, which will be fed back to the RL agent. These rewards can be accumulated during scheduling the whole sequence of jobs. In this study, we use minimizing the average bounded job slowdown ($bsld$) as the reward. The RL agent can learn from how its actions affect the future state and the corresponding rewards using the PPO algorithm. More details about each component are described in the next subsections.

\subsection{Observation}
The observation of RLBackfilling includes three parts: 1) the current job queue, 2) the selected job to run, and 3) the resource availability. For the `Current Job Queue', we will sort all the jobs based on their submission time when constructing the observation. For each job, we use its features to build a vector to represent it. For the `Selected Job', we treat it as a normal job in the queue. But we will introduce a mask to make sure the RL agent will never pick this job for backfilling. For the `Resource Availability', we assume the HPC environment is homogeneous in this study. Hence the availability is a percentage of available computing nodes for running jobs. Note that, instead of creating a separate scalar value with padding to align with job vectors, we simply append the resource availability into each job vector. In this way, each job vector will contain the resource availability information, which is the key for the kernel-based RL neural networks to work.

\subsection{RL Agent}
We use deep neural networks to implement the Backfilling Agent. Specifically, we follow the actor-critic model for better policy learning, using a policy and value network. 

\subsubsection{Policy Network.}
The policy network takes the observation as input and outputs an action for which job to run next. Similar to previous work~\cite{zhang2020rlscheduler}, we use a similar kernel-based neural network to improve performance and resolve potential order issues. The kernel network is insensitive to the order of jobs and uses a 3-layer fully connected network. The main difference is that our kernel-based network is applied to each job, calculating a score used to generate a vector. We then finally run softmax on the vector to generate a probability distribution for every waiting job.  

The output probability distribution serves two main functions. First, it is sampled during training to determine the next action. This makes the exploration of new actions/policies easier. Second, during testing, we directly select the job with the highest probability to ensure the best decision-making, as there is no more need for exploration. 
By using a kernel-based design, the parameter size of the policy network can be extremely small; as we only input a single job’s attributes each time.

\subsubsection{Value Network.}
The value network completes the actor-critic model, improving the efficiency of training. The architecture is a 3-layer MLP without the kernel mechanism, so the jobs are concat and flattened before being input into the network. 

The output of the value network is the expected reward $exp_r$ for a set of jobs based on the agent’s current backfilling policies. This indicates how well the agent can perform on the given set of jobs based on historical precedent. Rather than using the accumulated reward ($r$) directly, the policy network uses the $exp_r$ instead. The difference is that we are using the improvement of the current policy over historical policies for the set of jobs. This strategy reduces the variance of inputs for more efficient training. 

The policy and value network receive the same input, which includes queued jobs and resource availability. We use a vector to \(v_j\) to embed state info from each job, using multiple jobs to form a matrix. Each vector contains job attributes such as arrival time and request processors. The vector also includes available resources for the system. One issue we ran into with using DNNs to read waiting jobs is that the amount of waiting jobs may change, but our DNN only takes fixed-size vectors as input. To resolve this, we limit the program to only observe a fixed number (MAX\_OBSV\_SIZE) of jobs. If there are fewer than this amount, we pad the vector with zeros; if there are more, we selectively cut them off. This amount can be changed as it is a configurable training parameter, but it is by default 128. Many HPC job management systems like Slurm also limit pending jobs by the same order of magnitude. When cutting off extra jobs, we utilize FCFS (first come first serve) to sort all pending jobs and choose only the top (MAX\_OBSV\_SIZE) jobs.

\subsection{Actions, Rewards, and Environment}
For RLBackfilling, the actions are simply the selected jobs for backfilling. One key thing about backfilling is to make sure if a job is backfilled, it should not impact the start of the selected job. This is a rule that can be applied during heuristic backfilling. But we can not easily apply it during the RL agent makes the decisions, especially considering, at this moment, the scheduler does not know whether the decision will delay the selected job or not. Our solution in RLBackfilling is to introduce a large negative reward if such a requirement has been violated.   

Except in such an extreme case, rewards are just numerical feedback from the environment describing the effectiveness of the agent’s actions. In specific, the reward acts as a function in terms of average bounded slowdown. We define this function as {$reward = \frac{sjf - bsld}{sjf}$. In this regard, the reward is the percentage improvement in bsld score when compared to using FCFS as the base scheduler; and SJF for backfilling. This incentivizes the agent to maximize reward by minimizing the average bounded slowdown.}
Typically, the reward is collected at each step based on the agent's corresponding action. However, this is not the case for the average bounded slowdown, as the metric is dependent on the entire job sequence being scheduled before the calculation can be completed. Therefore, before the job sequence is complete, each step returns a reward of 0, only returning the true reward at the very last step. As only accumulated rewards are used for training, this does not reduce the effectiveness of the RL training process.


Training an RL model needs enormous interactions between the agent and the environment. It is impractical to perform such training in a real HPC cluster. 
Similar to previous work~\cite{Carastan-Santos2017ObtainingLearning,zhang2020rlscheduler}, RLBackfilling conducts training using a simulated HPC environment.
Specifically, we use an RL-compatible simulator implemented in RLScheduler~\cite{zhang2020rlscheduler}.  

\section{Evaluation}
\label{sec:eval}

This section evaluates RLBackfilling, with a particular emphasis on addressing the following research questions:
\begin{itemize}
    \item Does our reinforcement learning-based backfilling design outperform the EASY backfilling approach? If so, how significant is this improvement?
    
    \item Is the backfilling strategy we derived through reinforcement learning versatile enough to be compatible with various base scheduling policies or distinct workload patterns?
     
    \item Does the RL-driven backfilling strategy yield superior results compared to heuristic backfilling strategies based on Runtime Predictions?

\end{itemize}

These questions will be addressed in the subsequent evaluation subsections.

\subsection{Evaluation Setup}
\subsubsection{Implementation and Configurations}
We implemented RLBackfilling using the Proximal Policy Optimization (PPO) algorithm from OpenAI Spinning Up using PyTorch~\cite{schulman2017proximal}. 

The training for RLBackfilling is executed across multiple epochs. Within an epoch, jobs are sampled sequentially and scheduled using a 'Base Job Scheduling Policy' (e.g., FCFS or SJF), with the RL agent determining backfilling decisions. The scheduling for the whole job sequence yields an accumulated reward, forming a trajectory. After accumulating multiple trajectories, the agent's neural network undergoes training. Specifically, we gather 100 trajectories for each epoch, where each trajectory encompasses the scheduling decisions for 256 consecutive jobs. After each epoch, the RLBackfilling agent will be updated, undergoing 80 update iterations for both the policy and value networks. A learning rate of $10^{-3}$ is used and additional hyper-parameters will be detailed in the source code.

\subsubsection{Job Traces and Scheduling Policies.}

{We list the job traces used in the evaluations in Table~\ref{tab:jobtrace} and categorize them into two groups. The first group addresses}
the real-world traces from the SWF archive~\cite{Feitelson2014ExperienceArchive}. {The second group addresses} the synthetic traces generated based on a widely used workload model proposed in \cite{lublin2003}. We used different parameters in the model and generated two traces with different characteristics. 
As the sizes of these job traces are largely different, we leveraged the first 10K jobs from them in our evaluations.

\begin{table}[ht]
	\small
	\caption{List of job traces}
	\begin{center}
		\label{tab:jobtrace}
		\begin{tabular}{ccccccc}
			\toprule
			Name & Date &$size$ & $i_t$(sec) & $r_t$(sec) & $n_t$  & Runtime\\		
			\midrule
			SDSC-SP2 & 1998  & 128 & 1055 & 6687 & 11 & both\\
            HPC2N   & 2002 & 240 & 538 & 17024 & 6 & both\\
			\midrule
			Lublin-1 & -  & 256 & 771 & 4862 & 22 & AR\\
			Lublin-2 & -  & 256 & 460 & 1695 & 39 & AR\\
			\bottomrule
		\end{tabular}
	\end{center}
\end{table}

The key characteristics of these traces are also shown in the table, including the total number of processors in the cluster ($size$), average job arrival interval ($i_t$), average requested runtime ($r_t$), and average requested processors ($n_t$). In summary, the job traces are quite diverse in the presented characteristics. 

It is worth noting that the synthetic job traces (e.g., Lublin-1) do not have user-submitted Request Time. They only have the Actual Runtime (AR) to be used. For real-world traces (e.g., SDSC-SP2), both user-requested time and actual runtime information are available. 

\begin{table}[ht]
	\caption{List of schedulers}
	\label{tab:funcs}
	\begin{center}
		\begin{tabular}{cl}
			\toprule
			Name & priority function \\
			\midrule
			FCFS & $score(t) = s_t$\\
			SJF & $score(t) = r_t$\\
			WFP3 & $score(t) = -(w_t/r_t)^3*n_t$\\
			F1 & $score(t) = log_{10}(r_t)*n_t + 870*log_{10}(s_t)$\\
			\bottomrule
		\end{tabular}
	\end{center}
\end{table}

To evaluate the generality of RLBackfilling, we tested how it works with different base schedulers, including several heuristic schedulers and a state-of-the-art learning-based scheduler.
Table~\ref{tab:funcs} reports the priority functions used in these schedulers.
Here, FCFS schedules jobs in the same order as they were submitted (i.e., using $s_t$). 
SJF schedules jobs based on how long the job will run (i.e., using $r_t$). 
WFP3~\cite{tang2009fault} belongs to the scheduler family that combines multiple factors. More specifically, they favor jobs that have shorter run times, request fewer resources, and experience longer waiting times, representing expert knowledge in tweaking the priority functions. Scheduler F1 is the best scheduler selected from~\cite{Carastan-Santos2017ObtainingLearning}. It was built on brute force simulation and non-linear regression and represents the state-of-the-art batch job scheduler for the goal of minimizing \textit{average bounded slowdown}.

\subsection{RLBackfilling Training}
In this set of experiments, we first evaluate 1) whether the RL-based RLBackfilling can successfully learn the efficient backfilling strategy to gain better job execution performance and 2) how efficient and effective the learning is. 

\begin{figure}[ht!]
\begin{center}
    \includegraphics[width=0.48\textwidth]{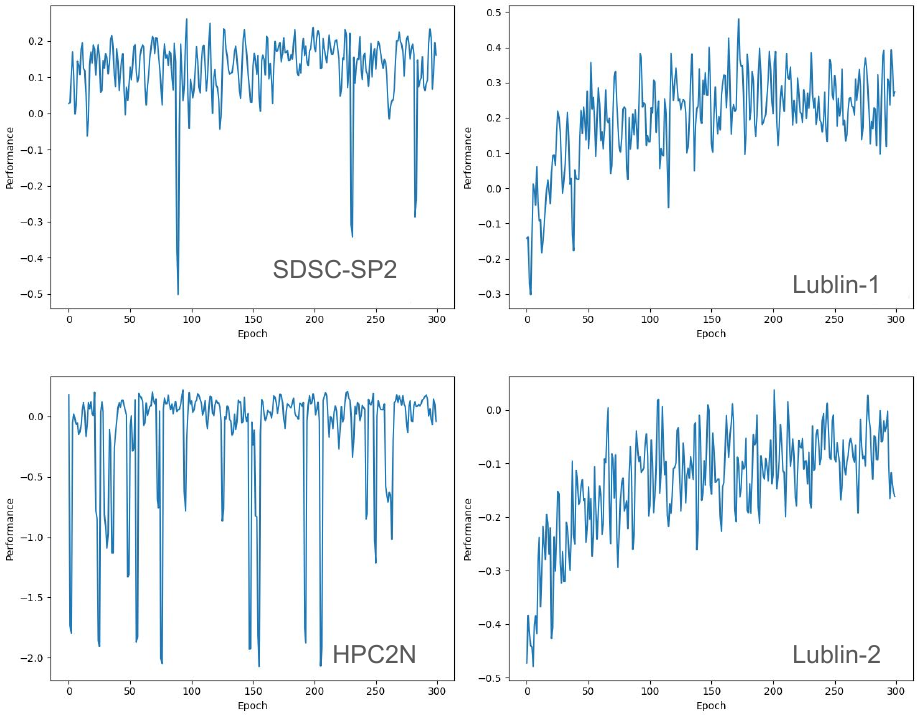}
	\caption{The training curves of RLBackfilling on four job traces (SDSC-SP2, HPC2N, Lublin-1, Lublin-2). \textit{x-axis} shows the training epoch; \textit{y-axis} shows the $bsld$ results.}
	\label{exp:curves}
\end{center}
\end{figure}

 To answer these questions, we directly show the training curves of RLBackfilling on four job traces listed in Table~\ref{tab:jobtrace} \textcolor{black}{using base scheduling policies FCFS (first come first serve). The optimization goal is the average bounded job slowdown percentage improvement relative to SJF backfilling.} Figure~\ref{exp:curves} presents the results on real-world and synthetic job traces. We discuss how RLBackfilling performs with other job scheduling policies and job execution performance metrics in detail in the next subsection.

 \textcolor{black}{The results in Figure~\ref{exp:curves} show RLBackfilling learns differently depending on the trace, with similar patterns in convergence with the synthetic and non-synthetic workloads. Though all of the models reached some amount of convergence, the non-synthetic workloads took noticeably longer. In contrast, we observe a much faster training process in the two synthetic workloads: Lublin-1 and Lublin-2, suggesting their regular job features and arrival patterns simplify the training procedure. On the other hand, training on HPC2N is less stable, frequently facing difficult trajectories of jobs. Despite this, the HPC2N model was to able to perform well on its own and other traces. These results use FCFS as the base scheduling policy.}



\subsection{RLBackfilling Scheduling Performance}

\textcolor{blue}{
\begin{table*}[ht!]
    \centering
    \caption{The actual performance of RLBackfilling when scheduling sampled job traces. \textit{RLBF indicates RLBackfilling.}}
    \begin{tabular}{|c|c|c|c|c|c|c|c|c|}
    \hline
        \textbf{Job Traces} & FCFS+EASY & FCFS+EASY-AR & \textbf{FCFS+RLBF} & SJF+EASY & SJF+EASY-AR & \textbf{SJF+RLBF} & WFP3+EASY & F1+EASY\\ \hline
        \hline
        \textit{SDSC-SP2} & 292.82 & 169.24 & 142.93 & 187.61 & 103.43 & 120.72 & 228.3 & 162.33 \\ \hline
        \textit{HPC2N} & 28.16 & 18.87 & 13.16 & 11.67 & 3.73 & 9.75 & 15.16 & 10.46 \\ \hline
        \textit{Lublin-1} & 192.89 & - & 83.43 & 55.62 & - & 30.57 & 138.89 & 50.9 \\ \hline
        \textit{Lublin-2} & 163.06 & - & 120.46 & 85.63 & - & 105.59 & 248.02 & 129.83 \\ \hline
    \end{tabular}
    \label{tab:1}
\end{table*}}

We further discuss the performance of RLBackfilling on actually backfilling jobs during scheduling different job traces. {Here, we use average bounded slowdown ($bsld$) itself as the metric.}

Note that, in these evaluations, we show the performance of RLBackfilling when the trained model is being used to schedule actual job sequences sampled from the same job trace (indicating the same workload pattern) but never seen in the training stage. Each time, we selected a random job sequence of 1024 jobs from the corresponding workloads and used the base scheduling policy (e.g., FCFS, SJF) plus different backfilling strategies (i.e., EASY and RLBackfilling) to schedule the same sequence and compare their performance. Note that, here we used much longer sequences ($1024$) than the one used for training ($256$). This was to benchmark if any overfitting occurred with the training dataset. We ran the evaluation 10 times each and reported the average for the sake of fair comparison. {Note that, we utilized different seeds for sampling the job traces to test each workload.}

The results are presented in Table~\ref{tab:1}. To facilitate analysis, we include several baselines for comparison. Specifically, for each job trace, we display results using two base scheduling policies (FCFS and SJF) combined with the EASY backfilling strategy. This EASY backfilling is implemented using the user-submitted Requested Runtime. To enable a comparison with Runtime Prediction-based strategies, we also incorporate results from base scheduling policies combined with EASY-AR. Here, EASY backfilling is carried out based on the Actual Job Runtime, which represents the ideal prediction. As we discussed earlier, the synthetic job traces such as Lublin-1 and Lublin-2 do not have user-submitted job runtime, so their results in EASY cases are omitted in the table. We also show the results of WFP3+EASY and F1+EASY as references to understand how the scheduling performance is impacted by the base scheduling policies.

From the results, we can conclude that RLBackfilling outperforms EASY in a variety of contexts. 
\textcolor{black}{First, it outperforms EASY based on user-submitted job runtime. 
For example, using base scheduling policy FCFS and RLBackfilling outperforms FCFS+EASY by 51\% $bsld$. Similar results can be observed in other job traces as well, with every model outperforming FCFS+EASY by at least 26\%.
More interestingly, it also outperforms EASY backfilling based on Actual Job Runtime, which indicates the perfect job runtime prediction results. For the SDSC-SP2 trace, the performance improvement is around 15\%. For HPC2N, FCFS+RLBF performs even better at 30\%.}These results confirm the benefits of using reinforcement learning to conduct backfilling as it can better learn the trade-off between prediction accuracy and backfilling opportunities.


Although the results are generated based on randomly sampling job traces that contain 1024 jobs (a much larger number than 256 jobs in the training set) to minimize the over-fitting, it is still a valid question to ask whether the RL agent is simply fitting the given trace instead of learning useful backfilling strategies. To validate that, in the next evaluations, we will further investigate RLBackfilling's generality towards unseen job traces. 

\subsection{RLBackfilling Generality}
As previously outlined, we further assessed RLBackfilling's performance when trained on one job trace, \textit{X}, and then applied to a distinct job trace, \textit{Y}. Given that the trained RL models have never encountered the applied trace, their performance can serve as a credible indicator of RLBackfilling's ability to learn effective and general strategies. This adaptability is crucial in real-world scenarios, where workload patterns consistently evolve.

\begin{table*}[ht]
    \caption{Performance comparisons of one RL-learned model (RL-$X$) being applied to other job traces (\textit{Y}).}
    \label{tab:cmp}
    \begin{center}
	\begin{tabular}{|c|c|c|c|c|c|c|}
    	\hline
    	\textit{Job Trace} & EASY & EASY-AR &RL-\textit{SDSC-SP2}      & RL-\textit{HPC2N} & RL-\textit{Lublin-1} & RL-\textit{Lublin-2} \\
    	\hline
		\multicolumn{7}{|c|}{\textit{FCFS as the Base Scheduling Policy}}\\
		\hline
		\textit{SDSC-SP2} & 292.82 & 169.24 & 142.93 & 187.18 & 216.94 & 210.09  \\ \hline
		\textit{HPC2N}    & 28.16  & 18.87  & 13.03 & 13.16  & 13.99  & 15.32           \\ \hline
		\textit{Lublin1}  & -      & 192.89 & 78.62  & 103.18 & 83.43  & 233.53  \\ \hline
		\textit{Lublin2}  & -      & 163.06 & 139.71 & 179.15 & 143.59 & 120.46  \\ \hline
		\midrule
		
		\hline
		\multicolumn{7}{|c|}{\textit{SJF as the Base Scheduling Policy}}\\
		\hline
		\textit{SDSC-SP2} & 187.61 & 103.43 & 120.72 & 125.52 & 160.53 & 133.17 \\ \hline
		\textit{HPC2N}    & 11.67  & 3.73   & 9.97  & 9.75   & 11.34  & 11.22           \\ \hline
		\textit{Lublin1}  & -      & 55.62  & 33.28  & 28.29  & 30.57  & 32.19   \\ \hline
		\textit{Lublin2}  & -      & 85.63  & 132.43 & 110.28 & 119.24 & 105.59  \\ \hline
		
        \bottomrule
	\end{tabular}
	\end{center}
\end{table*}

The results are shown in Table~\ref{tab:cmp}. Here, we separated the table into two sections. The top section shows the results of using FCFS as the Base Scheduling Policy. The bottom section uses SJF as the base scheduling policy. In each section, we show the results in multiple columns. The first column `EASY' show the $bsld$ results of using FCFS+EASY or SJF+EASY to schedule these job traces. The rest of the columns are named in the form of `RL-\textit{X}', which indicates the RLBackfilling was trained using job trace \textit{X}. For instance, RL-\textit{Lublin-1} indicates the results of reinforcement learning agent trained using \textit{Lublin-1} job trace. Each cell shows the performance of the corresponding RL model being applied to job trace \textit{Y}.

\textcolor{black}{From these results, we can have the following observations. First, the RL-based backfilling strategies are able to outperform EASY backfilling in most cases across all the job traces. For instance, the FCFS+EASY on HPC2N job trace yields $bsld$ result as $28.16$. However, the RLBackfilling model trained using HPC2N achieves much better performance ($bsld=13.16$). In addition, the RLBackfilling models trained using different job traces, such as SDSC-SP2, Lublin-1, or Lublin-2, also achieve much better performance than EASY backfilling. Even trained using a different job trace, the learned backfilling strategy is still applicable for other traces. Second, it is interesting to see that the RL model trained on \textit{X} (RL-\textit{X}) does not always perform the best in job trace \textit{X}. For instance, using FCFS as the base scheduling policy, we can see RL-\textit{SDSC-SP2} is able to outperform RL-\textit{Lublin-2} on its own trace; outperforming FCFS+EASY by 59\%. Despite facing a more difficult training workload, both of the models trained on non-synthetic data show generality in their backfilling performance.}

\section{Related Work}
\label{sec:related}
In the domain of High-Performance Computing, scheduling has always held paramount importance. One significant improvement in scheduling performance over the years can be attributed to backfilling~\cite{lifka1995anl,mu2001utilization,tsafrir2007backfilling,slack-basedbf,SelectiveReserv}. Various backfilling techniques such as Conservative~\cite{mu2001utilization}, Slack-Based~\cite{slack-basedbf}, and EASY backfilling~\cite{lifka1995anl} have been studied over time. 
Among them, EASY backfilling is the most popular and widely adopted one and has been used in mainstream resource managers such as Slurm~\cite{slurmweb}.
Existing backfilling techniques depend on either user-supplied runtime estimates~\cite{mu2001utilization}, which is too inaccurate and leads to suboptimal performance~\cite{tsafrir2007backfilling}, or runtime predictions~\cite{witt2019predictive,gaussier2015backfilling,fan2017trade,Tanash2019SupervisedRuntime,gaussier2015backfilling}, which believe better runtime prediction will lead to better scheduling performance. RLBackfilling is designed based on a different assumption. We observed better job runtime prediction might not always lead to better scheduling and designed RLBackfilling to directly make end-to-end backfilling decisions without explicitly predicting the job runtime, leveraging the reinforcement learning method. 


The potential of Reinforcement Learning in scheduling HPC jobs has been recently realized, with various schedulers incorporating RL to improve their performance, such as RLScheduler~\cite{zhang2020rlscheduler} , DRAS~\cite{Fan2021DeepRA}, RLSchert~\cite{RlSchert} and SchedInspector~\cite{zhang2022SchedInspector}. The studies of these schedulers indicate that RL can indeed yield superior scheduling results. However, our approach is unique as it is the first to apply RL specifically and systematically to backfilling. There are fundamental distinctions in realizing RL between scheduling and backfilling. For instance, while backfilling needs the consideration of relative jobs, this is not the case with generic scheduling. Additionally, in the context of backfilling, RL decision points occur at specific, distinct moments, whereas in scheduling, decision-making is a more regular and consistent process. Our approach RLBackfilling addresses these distinctions carefully and leads to different solutions and results.

\section{Conclusion and Future Work}
\label{sec:conclude}
{Our study proposes the integration of reinforcement learning (RL) into the backfilling process to enhance scheduling efficiency. Traditional backfilling algorithms such as EASY exhibited shortcomings due to their fixed nature and reliance on user-submitted job runtime information. Predicting job runtime has been taken but also exhibits problems as more accurate predictions do not necessarily indicate better scheduling performance. In contrast, our RLBackfilling approach showcased the potential for dynamic decision-making through adaptation based on learned patterns from past job trajectories. The effectiveness of RLBackfilling was extensively evaluated across various queue heuristics and workloads. Particularly compelling were the results when combining RLBackfilling with the SJF and FCFS queue heuristics, consistently outperforming EASY-Backfilling. This adaptability presented significant improvements, even though certain challenges like convergence issues for some job traces or certain unanswered questions like extremely good performance when training on some job traces were encountered and planned in our future work. In the broader landscape of high-performance computing, this research opens a promising way to easily integrate RL-based intelligent decision-making into existing HPC job scheduling, advancing computational performance in diverse application domains.}

\section*{Acknowledgments}
We sincerely thank the anonymous reviewers for their valuable feedback. This work was supported in part by NSF grants {CCF-1908843 and CNS-2008265}.

\bibliographystyle{ACM-Reference-Format}	
\bibliography{bib}
\end{document}